\documentclass[11pt]{article}
\usepackage{graphicx}
\usepackage{amssymb}
\usepackage{epstopdf}
\def\be{\begin{equation}}
\def\ee{\end{equation}}
\def\bea{\begin{eqnarray}}
\def\eea{\end{eqnarray}}
\def\nn{\nonumber}

\def\a{\alpha}
\def\b{\beta}

\def\d{\delta}
\def\e{\epsilon}
\def\f{\phi}
\def\g{\gamma}

\def\j{\psi}

\def\l{\lambda}
\def\m{\mu}

\def\o{\omega}

\def\q{\theta}

\def\s{\sigma}

\def\x{\xi}

\def\D{\Delta}

\def\G{\Gamma}

\def\L{\Lambda}
\def\O{\Omega}

\def\del{\partial}
\def\de{\nabla}
\def\mcal{\mathcal}

\def\ra{\rightarrow}

\DeclareSymbolFont{AMSa}{U}{msa}{m}{n}
\DeclareSymbolFont{AMSb}{U}{msb}{m}{n}
\DeclareMathSymbol{\fieldR}{\mathalpha}{AMSb}{"52}

\def\pp{{\mathchoice
          {
              \kern 1pt%
              \raise 1pt
              \vbox{\hrule width5pt height0.4pt depth0pt
                    \kern -2pt
                    \hbox{\kern 2.3pt
                          \vrule width0.4pt height6pt depth0pt
                          }
                    \kern -2pt
                    \hrule width5pt height0.4pt depth0pt}%
                    \kern 1pt
           }
            {
              \kern 1pt%
              \raise 1pt
              \vbox{\hrule width4.3pt height0.4pt depth0pt
                    \kern -1.8pt
                    \hbox{\kern 1.95pt
                          \vrule width0.4pt height5.4pt depth0pt
                          }
                    \kern -1.8pt
                    \hrule width4.3pt height0.4pt depth0pt}%
                    \kern 1pt
            }
            {
              \kern 0.5pt%
              \raise 1pt
              \vbox{\hrule width4.0pt height0.3pt depth0pt
                    \kern -1.9pt  
                    \hbox{\kern 1.85pt
 \vrule width0.3pt height5.7pt depth0pt
                          }
                    \kern -1.9pt
                    \hrule width4.0pt height0.3pt depth0pt}%
                    \kern 0.5pt
            }
           {
              \kern 0.5pt%
              \raise 1pt
              \vbox{\hrule width3.6pt height0.3pt depth0pt
                    \kern -1.5pt
                    \hbox{\kern 1.65pt
                          \vrule width0.3pt height4.5pt depth0pt
                          }
                    \kern -1.5pt
                    \hrule width3.6pt height0.3pt depth0pt}%
                    \kern 0.5pt
            }
        }}

 \def\mm{{\mathchoice
                       {
                             \kern 1pt
               \raise 1pt    \vbox{\hrule width5pt height0.4pt depth0pt
                                  \kern 2pt
                                  \hrule width5pt height0.4pt depth0pt}
                             \kern 1pt}
                       {
                            \kern 1pt
               \raise 1pt \vbox{\hrule width4.3pt height0.4pt depth0pt
                                  \kern 1.8pt
                                  \hrule width4.3pt height0.4pt depth0pt}
                             \kern 1pt}
                       {
                            \kern 0.5pt
               \raise 1pt
                            \vbox{\hrule width4.0pt height0.3pt depth0pt
                                  \kern 1.9pt
                                  \hrule width4.0pt height0.3pt depth0pt}
                            \kern 1pt}
                       {
                           \kern 0.5pt
             \raise 1pt  \vbox{\hrule width3.6pt height0.3pt depth0pt
                                  \kern 1.5pt
                                  \hrule width3.6pt height0.3pt depth0pt}
                           \kern 0.5pt}
}}
 \def\ad{{\kern0.5pt
                  \alpha \kern-5.05pt \raise5.8pt\hbox{$\textstyle.$}\kern
 0.5pt}}
 \def\bd{{\kern0.5pt
                   \beta \kern-5.05pt \raise5.8pt\hbox{$\textstyle.$}\kern
 0.5pt}}

 \def\qd{{\kern0.5pt
                   q \kern-5.05pt \raise5.8pt\hbox{$\textstyle.$}\kern
 0.5pt}}
 \def\Dot#1{{\kern0.5pt
     {#1} \kern-5.05pt \raise5.8pt\hbox{$\textstyle.$}\kern
 0.5pt}}

\begin{document}
\hbox to \hsize{{\tt arXiv:0707.1697 }\hss
\vtop{
\hbox{MCTP-07-19}
\hbox{UMDEPP-07-005}
\hbox{UK/07-03}
}}

\begin{center}
{\Large\bf  T-duality, quotients and generalized K\"ahler geometry }\\
[.5in]
\centerline{\large \bf Willie Merrell${}^{1,2}$ and Diana Vaman${}^{3}$  }

\vskip .5cm
\centerline{\it ${}^1$ Department of Physics}
\centerline{ \it University of Maryland}
\centerline{\it College Park, MD 20472}

\vskip .2cm
\centerline{\it ${}^2$ Department of Physics and Astronomy}
\centerline{ \it University of Kentucky}
\centerline{\it Lexington, KY 40506}

\vskip .2cm
\centerline{\it ${}^3$ Michigan Center for Theoretical
Physics}
\centerline{ \it Randall Laboratory of Physics, The University of
Michigan}
\centerline{\it Ann Arbor, MI 48109}

\vspace{1cm}

\begin{abstract}
In this paper we reopen the discussion of gauging 
the two-dimensional off-shell $(2,2)$ supersymmetric sigma models 
written in terms of semichiral superfields. 
The associated target space geometry of this particular sigma model is 
generalized K\"ahler (or bi-hermitean with two non-commuting complex 
structures).
The gauging of the isometries of the sigma model is now done by coupling the 
semichiral superfields to the new (2,2) semichiral vector multiplet. 
We show that the two moment maps together with a third function form the 
complete set of three Killing potentials which are associated with this 
gauging. We show that the Killing potentials lead to the generalized moment 
maps defined in the context of twisted generalized K\"ahler geometry.
Next we address the question of the T-duality map, 
while keeping the (2,2) supersymmetry 
manifest. Using the new vector superfield in constructing the duality 
functional, under T-duality we 
swap a pair of left and right semichiral superfields by a 
pair of chiral and twisted chiral multiplets.
We end with a discussion on quotient construction.
\end{abstract}
\end{center}
\newpage

\section{Introduction and Summary}

The geometry of the target space of two-dimensional sigma models is dictated by the amount of preserved world-sheet supersymmetry and by the representation of the sigma model fields. In the physics literature it has been known for quite a while \cite{ghr} that (2,2) supersymmetric sigma models give rise to special geometry manifolds. These are called bi-hermitean manifolds, and are endowed with a Riemannian metric $g$, a closed three-form $H=3 dB$, and two complex structures $J^{(\pm)}$. The metric is hermitean with respect to both complex structures, and $J^{(\pm)}$ are covariantly constant with respect to connections that have torsion determined by $H$.  More recently, it has been shown that the superfield representations needed for a complete description of the (2,2) supersymmetric sigma model include, beside the better known chiral and twisted chiral superfields, the semichiral superfields \cite{lruz2}. With only chiral and twisted chiral among the sigma model fields, th!
 e bi-hermitean geometry acquires an almost product structure, with the two complex structures commuting. In the case when the sigma model fields include the left and right semichiral superfields, the commutator of the two complex structures no longer vanishes. It is this latter case that we address in this paper.

In the mathematics literature, the study of the generalized Calabi-Yau manifolds, which include a non-trivial $B$-field, lead to the development of generalized complex geometry \cite{hit}. 
Its main object is the generalized complex structure
defined on the direct sum of the tangent and cotangent bundles $T\oplus T^*$. A special case of generalized complex geometry is the generalized K\"ahler geometry, which has two commuting generalized complex structures ${\cal J}_1,{\cal J}_2$, and a positive definite metric $G=-{\cal J}_1{\cal J}_2$.
Gualtieri \cite{thesis} has shown the equivalence of the data which define the bi-hermitean geometry with those of the generalized K\"ahler geometry. Therefore these two notions are interchangeable.
Special cases of the generalized K\"ahler geometry include symplectic and K\"ahler geometry. For recent work on related topics see \cite{long}.

An interesting question arises in the presence of isometries. In the symplectic case it is possible to talk about a Hamiltonian reduction, by defining the moment map (a function which is preserved by the action of the isometry group and follows from the requirement that the symplectic form is preserved) and restricting to the subspace of constant moment map. Similarly it is possible to define a K\"ahler quotient. The basic object associated with the quotient construction is the moment map. There are several proposals for the moment map definition in the context of generalized complex geometry. On the other hand, from the sigma model perspective, there is a natural place to look for the moment map, and that is the gauged sigma model. The moment map (sometimes referred to as Killing potential) appears in the off-shell (2,2) supersymmetric gauged sigma model action, multiplying the gauge superfield strengths. We are interested in matching the sigma model construction of the mom!
 ent map with the appropriate mathematical definition.

This paper is a follow up to \cite{Merrell:2006py}, and here we give the answer to the open questions of that previous work.  The new ingredient is the use of the appropriate (2,2) semichiral vector
multiplet \cite{Gates:2007ve, Lindstrom:2007vc} for the gauging of the (2,2) supersymmetric semichiral sigma model. This is the subject of Section 2.
We reduce the gauged action to (1,1) superspace. From the manifestly (1,1) supersymmetric gauged action we identify a set of three Killing potentials which multiply the various gauge superfield strengths.

In Section 3 we show how the Killing potentials relate to the moment maps.
We have done this by starting from the reduced (1,1) action, and inquiring what are the conditions which insure its invariance under the second set of non-manifest (1,1) supersymmetries. Besides the usual bi-hermitean geometry requirements, we found a set of conditions which express the two moment maps in terms of the Killing potentials. In the process we discovered that the third
Killing potential is instrumental in fixing a certain ambiguity in the definition of the moment maps (from general arguments, the moment maps are
defined up to a function $\sigma$ such that $d\sigma$ is invariant under the
action of the isometry group). After this ambiguity was fixed in the way we described, then  we were able to prove the existence of two generalized moment maps, defined for the twisted generalized K\"ahler structure in \cite{lin-tolman2}, one for each generalized complex structure.

In Section 4 we turn to the subject of T-duality.  Our starting point is the gauged (2,2) semichiral sigma model action. We construct the duality functional in (2,2) superspace.
Under T-duality, a pair of left and right semichiral superfields along the isometry direction and their antifields are replaced by a pair of chiral and twisted chiral superfields,
and their antifields. We work out the T-dual of the torus $T^4$ and show, at the level of (2,2) superspace, the usual map of the radius of the compact T-duality direction $R\leftrightarrow 1/R$.
Section 4.1 is dedicated to spelling out the role of the moment map in the
T-duality procedure.
Lastly, in section 5, we discuss the quotient construction, and work out one explicit example.
\section{The gauged (2,2) sigma model reduced to (1,1) superspace}
We begin by recalling the new gauged (2,2) supersymmetry algebra, which 
defines the new semichiral vector multiplet 
\cite{Gates:2007ve, Lindstrom:2007vc} (our notation follows 
\cite{Gates:2007ve}):
\bea
\label{algebra}
&&[\nabla_\a,\nabla_\b\}=4 \l (\g^3)_{\a\b}\bar T \x\cr
&&[\nabla_\a,\bar\nabla_\b\}=2i(\g^c)_{\a\b}\nabla_c-2\l
[iC_{\a\b}S+(\g^3)_{\a\b}P]\x\cr
&&[\nabla_\a,\nabla_b\}=-i\l(\g_b)_\a{}^\b \bar W_\b \x
+i\l(\g^3\g_b)_\a{}^\b\bar\O_\b \x\cr
&&[\nabla_a,\nabla_b\}=-\l\epsilon_{ab}\mcal{W}\x,
\eea
where the gauged supercovariant derivatives are defined as
\be
\nabla_\a=D_\a - \l\Gamma_\a \x.
\ee
The notation is such that $D_\a,\bar D_a$ are the usual (2,2) 
supercovariant derivatives, $\Gamma_\a$ is the superconnection, 
and $\x$ is the generator of the U(1) gauge transformation\footnote{Since we will be gauging a U(1) isometry of target space associated to a sigma model we have replaced the usual anti-hermitian U(1) generator denoted $t$ with $t = -i\x$, where $\x$ is the Killing vector for the isometry.}.
The associated Bianchi identities are:
\bea
\label{BI}
\nabla_\a S&=&-i\bar W_\a\cr
\cr
\nabla_\a P&=&-(\g^3)_\a{}^\b\bar W_\b\cr
\cr
\bar\nabla_\a T &=& 0\cr
\cr
\nabla_\a T &=&\O_\a \cr
\cr
\nabla_\a \O_\b &=& -C_{\a\b} \s \cr
\cr
\nabla_\a \bar\O_\b &=& 2i(\g^a)_{\a\b}\nabla_a \bar T \cr
\cr
\nabla_\a \bar W_\b&=&0\cr
\cr
\nabla_\a W_\b &=& iC_{\a\b}d-(\g^3)_{\a\b}(\s_1+\mathcal{W})+(\g^a)_{\a\b}\nabla_a S-i(\g^3\g^a)_{\a\b}\nabla_a P\cr
\cr
\nabla_\a d&=& (\g^a)_\a{}^\b\nabla_a \bar W_\b\cr
\cr
\nabla_\a \s &=& 0 \cr
\cr
\bar \nabla_\a \s &=& 2i(\g^a)_\a{}^\b\nabla_a\O_\b.
\eea
The constraints preserving the semichiral representation 
\be
(\gamma_a)^{\a\b}[\nabla_a,\nabla_\b\}=0
\ee
are solved by
\bea
\G_+=D_+\bar V_1, \qquad 
\G_-=D_-\bar V_2
\eea
and the standard constraints $(\gamma_a)^{\a\b} [\nabla_\a,\bar\nabla_\b\}=-4i\nabla_a$ 
allow for solving the vector superfield gauge potential, $\G_a$, 
in terms of the fermionic superfield gauge potential $\G_\a$.

It is perhaps useful to remind the reader that $\gamma^a=(\gamma^0,\gamma^1)$,
$(\gamma^0)_\alpha{}^\beta=\sigma_2, (\gamma^1)_\a{}^\b=i\sigma_1$, and that 
the indices are raised and lowered with $C_{\a\b}$ according to the 
north-west rule
\be
\gamma_{\a\b}=\gamma_\a{}^\delta C_{\delta\beta}, \gamma^{\a\b}=C^{\a\d}\gamma_\d{}^\b ,
\ee
where
\be
C^{\alpha\beta}C_{\gamma\beta}=\delta_\gamma^\alpha, 
\qquad C_{\alpha\beta}=\sigma_2.
\ee
To define our conventions more precisely, we write a two-component spinor as
$\theta^\a=(\theta^+,\theta^-)$. Alternatively, the $\pm$ indices denote the 
chiral components $\theta^\pm = \frac12 (1\pm\gamma^3)_\a{}^\beta 
\theta^\alpha$. Similarly, the for the derivatives $D_\pm$ we define
$D_\pm = \frac 12 (1\pm \gamma^3)_\a{}^\b D_\b$.

The gauge supercovariant algebra becomes
\bea
&&\{\nabla_+,\nabla_+\}=\{\nabla_-,\nabla_-\}=0,\qquad \{\nabla_+,\nabla_-\}
=-4i\l \bar T \xi,\nn\\
&&\{\nabla_+,\bar \nabla_-\}=2\l (-S+iP)\xi, \qquad \{\nabla_-,\bar\nabla_+\}=
2\l (S+iP)\xi,\nn\\
&&\{\nabla_+,\bar\nabla_+\}=2i\nabla_\pp, 
\qquad \{\nabla_-,\bar\nabla_-\}=2i\nabla_\mm,\nn\\
&&[\nabla_\pp,\nabla_\pp]=[\nabla_\mm,\nabla_\mm]=0,
\qquad [\nabla_\pp,\nabla_\mm]=-\l{\cal W}\xi,
\eea
where the bosonic gauge-covariant derivatives are denoted by $\nabla_\pp=2(\nabla_0+\nabla_1), \nabla_\mm=2(\nabla_0-\nabla_1)$.

As discussed in \cite{hullklr,Merrell:2006py}, the gauging of the sigma model can be done most straightforwardly at the level of (2,2) superspace. Here the sigma-model is defined entirely by the K\"ahler potential, which is a functional of the (2,2) superfields. 
The (2,2) superfields needed for a complete description of the two-dimensional
off-shell (2,2) supersymmetric sigma models are \cite{lruz2}:
\bea
&&{\rm chiral:}\qquad \bar D_\pm \phi=0, \qquad {\rm antichiral:}
\qquad D_\pm \bar \phi=0\nn\\
&&{\rm twisted \;chiral:} \qquad \bar D_+ \psi=D_-\psi=0,\qquad {\rm twisted 
\;antichiral:}\qquad  D_+\bar\psi=\bar D_-\bar \psi=0 \nn\\
&&{\rm left\; semichiral:} \qquad \bar D_+ X=0, \qquad {\rm left \; 
anti-semichiral:} \qquad D_+\bar X=0\nn\\
&&{\rm right\; semichiral:} \qquad \bar D_-Y =0, \qquad {\rm right  \; 
anti-semichiral:} \qquad D_-\bar Y=0.
\eea
In the case we are interested in, the K\"ahler potential depends on left and right semichiral superfields and their antifields\footnote{Both types of semichiral superfields are needed to define a sigma-model \cite{st}.}
\be
{\cal S}=\int d^2 \bar \theta d^2 \theta  K(X,Y,\bar X,\bar Y)
\ee
Next, one uses that the Grassmann integration is equivalent to   
differentiation. In order to couple the matter fields to the vector 
superfield, the supercovariant derivatives $D_\a,\bar D_\a$ are replaced by 
the gauged supercovariant derivatives $\nabla_\a,\bar\nabla_\a$. Lastly, we descend to the level of (1,1) superspace by replacing the (2,2) gauged supercovariant derivatives by two copies of (1,1) derivatives. The final step is to keep only one of the two (1,1) supersymmetries manifest, by reducing along the 
direction of the other (1,1). This will give the manifestly (1,1) supersymmetric gauged sigma model.

More concretely, the two (1,1) gauge supercovariant derivatives are defined by
\be
\hat\de_\a=\frac{1}{\sqrt{2}}(\de_\a+\bar\de_\a),~~~~\tilde\de_\a=\frac{i}{\sqrt{2}}(\de_\a-\bar\de_\a).
\ee
It is important to keep in mind that from the point of view of the 
(1,1) gauged sigma model, 
the $\tilde \nabla_\a$ derivatives act as the generators of the additional, 
non-manifest (1,1) supersymmetry. 

The $(1,1)$ gauge supercovariant derivatives obey the algebra
\bea
[\hat\de_\a,\hat\de_\b\}&=&2i(\g^a)_{\a\b}\de_a+2\l(\g^3)_{\a\b}(2T_1-P)\xi\cr
\cr
[\hat\de_\a,\de_b\}&=&-ig(\g_b)_\a{}^\b\hat W_\b \xi+i\l(\g^3\g_b)_\a{}^\b \hat\O_\b \xi\cr
\cr
\cr
[\tilde\de_\a,\tilde\de_\b\}&=&2i(\g^a)_{\a\b}\de_a-2\l(\g^3)_{\a\b}(2T_1+P)\xi\cr
\cr
[\tilde\de_\a,\de_b\}&=&-ig(\g_b)_\a{}^\b\tilde W_\b \xi+i\l(\g^3\g_b)_\a{}^\b \tilde\O_\b \xi\cr
\cr
\cr
[\hat\de_\a,\tilde\de_\b\}&=&4\l(\g^3)_{\a\b}T_2\x-2\l C_{\a\b}S\x\cr
\cr
[\de_a,\de_b\}&=&-\l\e_{ab}\mcal{W}\xi
\eea

The $(2,2)$ fermionic measure is evaluated using the $(2,2)$ gauge 
supercovariant derivatives
\be
\label{measure}
\int d^2\bar\q d^2\q=\frac{1}{8}[\nabla^\a\nabla_\a\bar\nabla^\b\bar\nabla_\b+
\bar\nabla^\b\bar\nabla_\b\nabla^\a\nabla_\a].
\ee
Using the relation
\be
\de_\a \de_\b \de_\g=\frac{2}{3!}\l\bar T (\g^3)_{(\a\b}\de_{\g)}\xi-\frac{8}{3!}\l C_{\b(\a}(\g^3)_{\g)}{}^\d \bar T \de_\d \xi,
\ee
we can show that
\be
\hat\nabla^\a\hat\nabla_\a\tilde\nabla^\b\tilde\nabla_\b=2\nabla^\a\nabla_\a
\bar\nabla^\b\bar\nabla_\b+2\bar\nabla^\b\bar\nabla_\b\nabla^\a\nabla_\a
+(...) \xi+\mbox{total derivative}.
\ee
This allows us to rewrite the fermionic measure in terms of the $(1,1)$ derivatives as
\be
\int d^2\bar\q d^2\q=\frac{1}{16}\hat\nabla^\a\hat\nabla_\a\tilde\nabla^\b\tilde\nabla_\b.
\ee
The implicit assumption here is that the K\"ahler 
potential that we are gauging is invariant under the symmetry transformation, 
i.e. it satisfies $\xi K=0$ (there is of course the possibility that the K\"ahler potential is invariant up to general K\"ahler transformations; the extension to this case, though relatively straightforward, is not addressed in this paper).

We now reduce the manifestly (2,2) supersymmetric action to $(1,1)$ superspace by evaluating the $(1,1)$ derivatives $\tilde\de^\a \tilde\de_\a$ onto the K\"ahler potential. After some algebra, we obtain
\bea
\label{reduction}
\tilde\nabla^\a\tilde\nabla_\a K&=&\frac{i}{2}[\hat\nabla_+
\varphi^Im_{II'}\hat\nabla_-\chi^{I'}
+\Upsilon_+^{I'}n_{I'I}\Psi_-^I
+\Psi_-^I(2\o_{IJ}\hat\nabla_+\varphi^J+ip_{II'}\hat\nabla_+\chi^{I'})\cr
&~&+\Upsilon_+^{I'}(2\o_{I'J'}\hat\nabla_-\chi^{J'}-iq_{I'I}\hat\nabla_
-\varphi^I)]\cr
&~&+8i\l [K_i(\xi X^i)-K_{\bar i}(\xi \bar X^{\bar i})-K_{i'}(\xi Y^{i'})+K_{\bar i'}(\xi \bar Y^{\bar i'})]T_2\cr
&~&-4i\l [K_i(\xi X^i)-K_{\bar i}(\xi \bar X^{\bar i})+K_{i'}(\xi Y^{i'})-K_{\bar i'}(\xi \bar Y^{\bar i'})]S\cr
&~&+2\l [K_i(\xi X^i)+K_{\bar i}(\xi \bar X^{\bar i})-K_{i'}(\xi Y^{i'})-K_{\bar i'}(\xi \bar Y^{\bar i'})](2T_1+P).
\eea
where we have kept the notation of \cite{lruz}: $I=(i,\bar i), \varphi=X|, \chi=Y|,$ etc...
By inspecting the resulting (1,1) sigma model action, we see that, as 
expected, we have the same metric and NS-NS two-form obtained in 
\cite{lruz}. 
However, there are some differences with respect to the case 
when the gauging of the U(1) isometry is done by using the usual (2,2) super 
Yang-Mills multiplet \cite{Merrell:2006py}. 
These differences are visible in the terms which 
depend on the superfield strengths. We shall focus on this aspect in the next 
section.

\section{Moment maps}

In the case of K\"ahler geometry, which is the target space geometry 
associated with a sigma model derived from a (2,2) chiral superfield-dependent 
K\"ahler potential \cite{zumino}, the gauging of an isometry requires that the generator
of the isometry preserves not only the metric (i.e. it is Killing) but the 
complex structure as well (i.e. it is holomorphic). 
As a consequence, the isometry generator preserves the symplectic form
$\omega=gJ$. Therefore, 
\be
{\cal L}_\xi \omega= i_\xi d\omega + d(i_\xi\omega)=0
\ee
implies that $i_\xi \omega$ is locally exact. This defines  the moment map
\be
i_\xi \omega= d\mu,
\ee
also referred to as the Hamiltonian function for symplectic manifolds, and 
the Killing potential for K\"ahler manifolds \cite{hklr}.
In the latter case, by going to the holomorphic 
coordinate base which diagonalizes the 
complex structure, and using that 
$\omega=2i \partial_{i\bar j}K d\phi^i\wedge d\bar \phi^{\bar j}$,  one finds
\be
-i \xi^i \partial_{i\bar j}K=\partial_{\bar j} \mu, \qquad 
i \xi^{\bar j}\partial_{i\bar j} K=\partial_i \mu.
\ee
This can be integrated in the case of an U(1) isometry to yield
\be
\mu=-i \xi^i \partial_i K+i \xi^{\bar j}\partial_{\bar j}K,
\ee 
up to a constant.

Studying (2,2) supersymmetric two-dimensional sigma models, 
Gates, Hull and Rocek  \cite{ghr} 
showed that their target space admits a bi-hermitean 
metric (hermitean with respect to two complex structures). The complex
structures are covariantly constant with respect to a torsion-full connection.
The torsion is related to the field strength of a two-form potential, the 
$B$ field. 
In the mathematics literature, the bi-hermitean geometry is known as generalized 
K\"ahler geometry \cite{thesis}.

If the (2,2) supersymmetric sigma model employs only chiral and twisted chiral
superfields, the two complex structures commute. This type of geometry is
referred to as an almost product structure space \cite{ghr}.  
As in the previous case, the
moment map follows from requiring that the isometry generator preserve the 
anti-symmetric two-forms $\omega^{(\pm)}=gJ^{(\pm)}$. 
This means that 
\be
{\cal L}_\xi \omega^{(\pm)} = 0.
\label{ham}
\ee
In the case of generalized K\"ahler geometry, $\omega^{(\pm)}$ is no longer 
a closed form, rather in the presence of a non-trivial $B$-field it satisfies
\be
\pm d\omega^{(\pm)} (J^{(\pm)} X,J^{(\pm)} Y,J^{(\pm)} Z)=dB (X,Y,Z). 
\ee
Then from (\ref{ham}) it follows that
\be
d\mu_\pm = \omega^{(\pm)} \cdot \xi \mp J^{(\pm)}{}^T \cdot u, 
\label{mom_map}
\ee 
where
\be
i_\xi H= du\label{u},\qquad H=3 dB.
\ee
When $\mu_\pm$ can be defined globally they are called moment maps.
Since the isometry generator $\xi$ preserves the complex structures, it 
respects the natural decomposition of the tangent space induced by the 
chiral $\phi^i$ and twisted chiral $\psi^{i'}$ coordinates. 
For $\xi=\xi^i \partial_i + \xi^{\bar i} \partial_{\bar i}$, the 
gauging of the sigma model is done by coupling with an ordinary (2,2) vector 
multiplet. For 
$\tilde \xi=\tilde \xi^{i'}\partial_{i'}+\tilde 
\xi^{\bar i'}\partial_{\bar i'}$,
the gauging is done by coupling with a twisted (2,2) vector multiplet 
\cite{hps}
\footnote{The large vector multiplet introduced in \cite{Lindstrom:2007vc}   
can be used to gauge an isometry which mixes the chiral and twisted chiral directions.
}. 
Following an off-shell (2,2) supersymmetric sigma model analysis, 
Hull, Papdopoulos and Spence \cite{hps} showed that 
the moment maps are identified with the Killing potentials $i\xi^{\bar i}
\partial_{\bar i} K$ and 
respectively $i\tilde \xi^{\bar i'}\partial_{\bar i'} K$.
In terms of the significance of the moment maps for the generalized K\"ahler 
geometry, it can be shown that is either the sum 
or the difference of the two moment maps $\mu_+, \mu_-$ which defines an 
eigenvector of the generalized complex structure ${\cal J}_{1/2}$ \cite{Merrell:2006py}, i.e.
$(\xi \pm \frac i2 (d\mu_+ \pm d\mu_-))\in T\oplus T^*$ lies in the eigenbundle 
of ${\cal J}_{1/2}$.

Lastly, we turn to the generic case of bi-hermitean geometry 
with non-commuting complex structure, which is realized 
by a semichiral superfield sigma-model \cite{lruz}.

In \cite{Merrell:2006py} it was found by studying a certain example of generalized K\"ahler
geometry, the $SU(2)\times U(1)$ WZNW sigma model, that the two a priori 
distinct moment maps are indeed distinct. This point deserves a further 
clarification since the on-shell (2,2) supersymmetric sigma model analysis 
in \cite{Kapustin:2006ic} points out to the existence of a unique moment map, 
with $\mu_+$ and $\mu_-$ being identified. 
In this paper we extend the investigation opened in \cite{Merrell:2006py} of an off-shell 
supersymmetric gauged (2,2) sigma model, by appropriately coupling the 
semichiral superfields with the newly found (2,2) semichiral vector 
multiplet \cite{Gates:2007ve, Lindstrom:2007vc}. 
In the process we shall find that besides 
the two moment maps there is a third function, called $\sigma$ in \cite{Merrell:2006py}, 
which together with the two distinct moment maps forms the complete set of 
three Killing potentials.  

The connection between the moment maps and the gauged sigma model action was 
previously discussed in \cite{hps}.
The idea is to start from the reduced (1,1) supersymmetric sigma-model action, 
and require that it is invariant under the additional, non-manifest (1,1) 
supersymmetries generated by $\tilde\nabla_\pm$.
These act on the (1,1) sigma-model superfields as
\bea
\delta \Phi&=&\frac{i}{\sqrt 2}\bigg[\epsilon^+
(\nabla_+-\bar\nabla_+)+\epsilon^-(\nabla_--\bar\nabla_-)\bigg]\Phi
\nn\\&=&\frac{1}{\sqrt 2}\bigg(
\epsilon^+ J^{(+)} \cdot \hat\nabla_+\Phi + \epsilon^- J^{(-)}\cdot 
\hat\nabla_-\Phi \label{susy1}
\bigg),
\eea
where $\Phi$ stands for the sigma-model superfields 
$\varphi^I,\chi^{I'}$ \cite{lruz}.

The action of the non-manifest supersymmetries on the gauge superconnections is inferred from:
\be
\delta \hat\nabla_\pm \Phi^m=\pm 2i\l \epsilon^\mp (S\pm 2T_2) \xi^m
-\epsilon^+\hat\nabla_\pm (J^{(\pm)}{}^m{}_n\hat\nabla_\pm\Phi^n)
-\epsilon^-\hat\nabla_\pm (J^{(\mp)}{}^m{}_n\hat\nabla_\mp\Phi^n)\label{susy2}.
\ee
Further using that $S-iP$ is a twisted chiral superfield and that $T$ is chiral, we find the non-manifest supersymmetry variation of the field strength superfields:
\bea
&&\delta(S-iP)={i}\bigg(-\epsilon^+ \hat \nabla_+ + 
\epsilon^- \hat \nabla_-\bigg) (S-iP) 
\nonumber\\
&&\delta T={i}\bigg(\epsilon^+\hat \nabla_+ + \epsilon^- \hat \nabla_-\bigg)T
\label{susy3}.
\eea

Let us now concentrate on the invariance of the manifestly (1,1) 
supersymmetric gauged sigma model action
\be
{\cal S}=\int d^2 x d^2 \hat\theta \bigg(2i\hat \nabla_+ \Phi \cdot (g+B) \cdot \hat\nabla_-\Phi
+ 4\l S\mu_1 -8\l T_2 \mu_2 + 2\l \sigma (2T_1+P) \bigg)
\label{gauged_action}
\ee
under the additional (\ref{susy1},\ref{susy2},\ref{susy3}) supersymmetries.
In the case we are investigating we have assumed that the K\"ahler potential 
is strictly invariant under the action of the U(1) isometry generator $\xi$. Because of this assumption, the first term in the gauged action is actually obtained by minimal coupling. In other words, since ${\cal L}_\xi g=
{\cal L}_\xi B=0$, then the kinetic terms and the B-field dependent terms in the sigma-model are gauged in the same way, by minimal coupling.  
We have introduced the notation $\mu_1,\mu_2$ for the 
terms which multiply the superfield strengths $S, T_2$ in (\ref{gauged_action}), even though we have their concrete expression in terms of derivatives of the 
K\"ahler potential from  (\ref{reduction}).
The reason for our feigned ignorance is that we want to be able to show the rapport between $\mu_1,\mu_2$ and the moment maps. This will become transparent once 
we require that (\ref{gauged_action}) has the additional (1,1) supersymmetries.

The invariance of (\ref{gauged_action}) is conditioned, among other things, 
by the cancellation of the terms in $\delta {\cal S}$ which are proportional 
to the superfield strengths $S, P, T_1, T_2$.
Those terms which are proportional to $S$ are
\bea
&&
4\l \epsilon^+ \bigg(-
\xi^m (g+B)_{nm} + \partial_m \mu_1 J^{(+)}{}^m{}_n-\frac 12 
\partial_n\sigma \bigg) \hat\nabla_+\Phi^n
\nonumber\\
&+&4\l \epsilon^- \bigg(-\xi^m (g+B)_{mn} + \partial_m \mu_1 J^{(-)}{}^m{}_n
+\frac 12 \partial_n\sigma\bigg) \hat \nabla_- \Phi^n.
\eea
Therefore we find that
\be
d\mu_1= -\xi \cdot (g-B)\cdot J^{(+)}-\frac 12 d\sigma \cdot J^{(+)},
\qquad
d\mu_1= -\xi \cdot (g+B)\cdot J^{(-)} + \frac 12 d\sigma \cdot J^{(-)}
\label{constr1}.
\ee
Similarly, the terms which are which are proportional to $T_2$ are
\bea&&8\l \epsilon^+\bigg(-\xi^m (g+B)_{nm} - \partial_m \mu_2 J^{(+)}{}^m{}_n
+\frac 12 \partial_n\sigma \bigg)\hat\nabla_+\Phi^n
\nonumber\\
&+&8\l\epsilon^-\bigg(\xi^m (g+B)_{mn}-\partial_m\mu_2 J^{(-)}{}^m{}_n +
\frac 12 \partial_n\sigma \bigg)\hat\nabla_-\Phi^n,
\eea
which implies that the action is invariant provided that
\be
d\mu_2=\xi\cdot (g-B)\cdot J^{(+)}-\frac 12 d\sigma \cdot J^{(+)},\qquad
d\mu_2=-\xi\cdot (g+B) \cdot J^{(-)}-\frac 12 d\sigma \cdot J^{(-)}
\label{constr2}.
\ee
In order for these two sets of equations to be satisfied, $\sigma$ must be such that 
\be
d\sigma=(d\mu_1+d\mu_2)\cdot J^{(+)} =-(d\mu_1-d\mu_2)\cdot J^{(-)}.
\ee 

To complete our investigation of the relationship 
between the Killing potentials $\mu_1,\mu_2, \sigma$ and the moment maps 
$\mu_+,\mu_-$, we recall that we have worked under the assumption that
the K\"ahler potential is invariant under the action of the isometry generator
$\xi K=0$. With the metric and $B$-field determined by the 
invariant K\"ahler potential, then ${\cal L}_\xi g={\cal L}_\xi B=0$. As a consequence, the one form 
$u$ defined in (\ref{u}) can be explicitly solved
\be
{\cal L}_\xi B=d(i_\xi B)+i_\xi dB=0\qquad \Rightarrow \qquad 
u=-\xi \cdot B+\tilde \sigma,
\ee 
where $d\tilde \sigma$ is an exact one-form, invariant under $\xi$. 
What (\ref{constr1}) and (\ref{constr2}) show is that 
\be
u=-\xi\cdot B+ \frac 12 \sigma
\ee
and that $2\mu_1$ and $2\mu_2$ are equal to the sum and respectively the 
difference of the moment maps $\mu_\pm$.  

The supersymmetry variations which are proportional to the superfield 
strengths $P$ and $T_1$ give rise to an equivalent set of constraints.
There are three terms which are proportional to each of these 
superfield strengths.
Two of these terms are obvious, coming from supersymmetry variations of 
$(\delta S) \mu_1$ and $P(\delta \sigma)$, and similarly for the terms proportional to $T_1$.
The third term will arise from the supersymmetry variations of $\hat\nabla_-\Phi \cdot
(g+B)\cdot \hat\nabla_+\Phi$, where we keep the contributions coming from the 
second and third term in (\ref{susy2}). After partial integration, these 
terms combine by using the anticommutator $\{\hat \nabla_+,\hat\nabla_-\}$. 

Of course, in addition to these constraints, in order to ensure the 
invariance of the action under the non-manifest (1,1) supersymmetries, 
the metric and $B$ field must satisfy the usual requirements
which define the bi-hermitean geometry. 
A perhaps unexpected requirement emerging from our supersymmetry 
analysis is that $E=g+B$ 
ought to be bi-hermitean. 
This is, however, in complete agreement with the manifestly (2,2) 
supersymmetric origin of the (1,1) action (\ref{gauged_action}). From a 
(2,2) superspace perspective, the complex structures, 
the metric and the $B$ field arise from second 
order derivatives of the K\"ahler potential \cite{lruz}. These explicit 
expressions enable the check that indeed $E=g+B$ is bi-hermitean.  These expressions should also allow a demonstration that the constraints, equations (\ref{constr1}, \ref{constr2}), are also satisfied.  While we were unable to show this in general, we have observed that they hold in the flat space and $SU(2)\otimes U(1)$ examples.

\subsection{Generalized Moment Maps}

In \cite{Merrell:2006py, Kapustin:2006ic} an effort was made to check whether the moment maps obtained from the sigma model correspond to the moment maps used in \cite{lin-tolman, lin-tolman2} as part of the definition of generalized moment maps.  The equations derived at the end of Section 3 allow us to extend these previous attempts to (2,2) supersymmetric sigma models with semichiral superfields, i.e. sigma models with three Killing potentials: the two moment maps and the function $\s$.  More explicitly, the equations (\ref{constr1}) and (\ref{constr2}) 
can be rewritten as
\bea
2d\m_1 &=& (\o^{(+)} + \o^{(-)})\xi -(J^{(+)}{}^T - J^{(-)}{}^T)(-\xi B+
\frac{1}{2}d\s)\cr
\cr
2d\m_2 &=& (\o^{(+)} - \o^{(-)})\xi +(J^{(+)}{}^T + J^{(-)}{}^T)(-\xi B - 
\frac{1}{2}d\s)\cr
\cr
2d\m_1 &=& -(J^{(+)}{}^T - J^{(-)}{}^T)d\s\cr
\cr
2d\m_2 &=& -(J^{(+)}{}^T + J^{(-)}{}^T)d\s~~.
\eea
{}From these equations it follows
\bea
\label{fav}
0 &=& (J^{(+)} - J^{(-)})\xi - (\o^{(+)}{}^{-1}+\o^{(-)}{}^{-1})u\cr
\cr
2d\mu_1 &=& (\o^{(+)} + \o^{(-)})\xi -(J^{(+)}{}^T - J^{(-)}{}^T)u~~,
\label{mu1eqn}
\eea
where we have used that $u=-\xi B+\frac{1}{2}d\s$.
As in \cite{Kapustin:2006ic}, (\ref{mu1eqn}) can be written in terms of 
${\cal J}_2$, one of the generalized complex structures given in \cite{thesis}, to show that that $\xi+u-id\m_1$ is an eigenvector of $\mcal{J}_2$.  This corresponds to the definition of a generalized moment map for twisted generalized K\"ahler geometry \cite{lin-tolman2}.  The proof is given by noting that since
\be
\mcal{J}_2=\frac{1}{2}\left(\begin{array}{cc}
J_+ - J_-& -(\o_+^{-1}+\o_-^{-1})\\
\o_+ + \o_- & -(J_+^t - J_-^t)\\
\end{array}\right),
\ee
then (\ref{fav}) can be written as $\mcal{J}_2(\xi+u)=d\m_1$.  This is equivalent to the equation $\mcal{J}_2(\xi+u-id\m_1)=i(\xi+u-id\m_1)$ which verifies the claim.  Similarly, $d\m_2$ is 
used in the construction of a second twisted generalized moment map, eigenvector of $\mcal{J}_1$.
\section{T-duality}

Next we discuss T-duality.  We follow the basic procedure outlined in 
\cite{rv}. 
First we gauge the U(1) isometry of the sigma model using the prepotentials 
of the gauge multiplet. 
Then we add the Lagrange multipliers which will force the 
field strength of the gauge multiplet to vanish.  
In the last step leading to the duality functional, we use the gauge freedom 
to gauge away the appropriate superfields.  
By solving the Lagrange multiplier constraints and substituting back into the 
duality functional we return to the original action.  The dual action is  
obtained by imposing the prepotential equations of motion.  We will 
work out one concrete example, $T^4$,  and observe 
the characteristic interchange of the $S^1$ radius $R\leftrightarrow 1/R$ 
in the T-dual actions.

In our discussion of T-duality we will maintain manifest the (2,2) 
supersymmetry. The gauging of the (2,2) supersymmetric sigma model action \cite{Gates:2007ve} is 
done at the level of the K\"ahler potential by replacing the left 
and right semichiral superfields $X$ and $Y$ by: 
\be
X\ra \tilde X=e^{V_1\xi}X, ~~~~\bar X\ra \bar{\tilde X}=e^{\bar V_1\xi}
\bar X,~~~~Y\ra \tilde Y=e^{V_2\xi}Y, \bar Y\ra \bar{\tilde Y}=
e^{\bar V_2\xi}\bar Y
\label{replacement},
\ee
where $V_1$ and $V_2$ are the prepotentials of the semichiral vector multiplet.  The prepotentials are worth a brief review as their transformations are important for the discussion of T-duality and quotients.  The form of the gauge covariant derivative algebra requires that the fermionic gauge potentials satisfy $\G_+ = D_+ \bar V_1$ and $\G_- = D_- \bar V_2$.  Since the gauge covariant derivatives are invariant under $\d\G_\a = D_\a L$, the prepotentials share a common transformation by a real scalar superfield denoted $L$,
\be
\label{Ltrans}
\d_L V_1 = L,~~ \d_L V_2 = L.
\ee
One can also note, that since left semichiral superfields are in the kernel of $\bar D_+$ and right semichiral superfields are in the kernel of $\bar D_-$ that $V_1$ and $V_2$ can transform by left and right semichiral superfields respectively, i.e.
\be
\label{semitrans}
\d_\L V_1 = \L,~~ \d_U V_2 = U,
\ee
where $\L$ is left semichiral and $U$ is right semichiral.  The substitutions given in (\ref{replacement}) as part of the gauging prescription replace the field $X$, which satisfies the regular semichiral constraint $\bar D_+ X = 0$, with $\tilde X$, which satisfies the gauge covariant semichiral constraint $\nabla_+\tilde X = 0$, and likewise for $Y$.  The covariantly semichiral superfields $\tilde X$ and $\tilde Y$ as well as the covariantly semichiral constraints are then invariant with respect to (\ref{semitrans}) and transform covariantly with respect to (\ref{Ltrans}). Here we assume that the gauging is done such that the K\"ahler potential is left unchanged under a gauge transformation (in a more general case, the 
K\"ahler potential may change by a general K\"ahler transformation, which
leaves the sigma model metric invariant).  The replacement in (\ref{replacement}) ensures the invariance of the 
gauged K\"ahler potential
\be
K_g=K(\bar{\tilde X},\tilde X,\bar{\tilde X},\tilde Y).
\ee
Because the isometry being gauged is a U(1), we can realize it as a shift in 
the superfields. The K\"ahler potential dependence on the fields is
\bea
K&=&K(X+Y,\bar X +\bar Y,\bar X+X,\bar Y+Y)
\cr
&&\mbox{or}\cr
K&=&K(\bar X+Y,X +\bar Y,\bar X+X,\bar Y+Y).
\eea
This would result in the gauge potentials
\bea
K_g&=&K(X+Y+i(V_1-V_2),\bar X +\bar Y+i(\bar V_2-\bar V_1),\bar X+X+i(V_1-\bar V_1),\bar Y+Y-i(V_2-\bar V_2))\cr
&&\mbox{or}\cr
K_g&=&K(\bar X+Y+i(V_2-\bar V_1),X +\bar Y+i(V_1-\bar V_2),\bar X+X+i(V_1-\bar V_1),\bar Y+Y-i(V_2-\bar V_2)).\cr
&~&
\eea
For concreteness, let's assume that the U(1) generator is 
\be
\xi=\partial_X-\partial_{\bar X}-\partial_Y+\partial_{\bar Y}
\ee
and accordingly, 
\be
K\equiv K (X+Y,X+\bar X,Y + \bar Y)=K(X+Y-\bar X-\bar Y, X+\bar X, Y+\bar Y).
\ee
On the tangent bundle $T$ we can define 
three other vectors, which together with with  $\xi$ form a vector 
basis:
\be
\xi_1 = \partial_X+\partial_{\bar X},\qquad \xi_2 = 
\partial_{Y}+\partial_{\bar Y}, \qquad \xi_3 =\partial_{X}+\partial_{Y}-\partial_{\bar X}-\partial_{\bar Y}.
\ee
The gauged K\"ahler potential can then be rewritten as
\bea
K_g&=& K + \frac{e^{2 i Im V_1 \xi_1} -1}{Im V_1 \xi_1} Im V_1 \xi_1 K + 
\frac{e^{2i  Im V_2 \xi_2}-1}{Im V_2 \xi_2} Im V_2 \xi_2 K \nn\\&+& 
\frac{e^{2 (Re V_1 - Re V_2)\xi_3}-1}{(Re V_1 - Re V_2)\xi_3} 
(Re V_1 - Re V_2)\xi_3 K\nn\\
&=& K + \frac{e^{2 i Im V_1 \xi_1} -1}{Im V_1 \xi_1} Im V_1 (\mu_1+\mu_2)
+\frac{e^{2 i Im V_2 \xi_2}-1}{Im V_2 \xi_2} Im V_2  (\mu_1-\mu_2)\nn\\ 
&+& 
\frac{e^{2 (Re V_1- Re V_2)\xi_3}-1}{(Re V_1 -Re  V_2)\xi_3} 
(Re V_1- Re V_2)\sigma\label{kgmu},
\eea
which emphasizes the role of $\mu_1,\mu_2$ and $\sigma$ as Killing potentials.
In addition, this  expression of the gauged K\"ahler potential makes 
manifest the dependence on only three of the four prepotentials, 
besides the three Killing potentials.

In order to select the appropriate  supersymmetry representation 
of the Lagrange 
multipliers, we first solve  the gauge superfield strengths in terms of 
the prepotentials
\bea
T&=&\frac{1}{4}\bar D^2(V_2-V_1)\cr
\bar T&=&\frac{1}{4}D^2(\bar V_2-\bar V_1)\cr
S+iP&=&\frac{1}{2}D_-\bar D_+(\bar V_2-V_1)\cr
S-iP&=&\frac{1}{2}D_+\bar D_-(V_2-\bar V_1).
\eea
As mentioned in the previous section, $T$ is a chiral superfield and $S-iP$ 
is twisted chiral.

Next we add the Lagrange multipliers which enforce the condition that 
the gauge field is pure gauge (i.e. its field strength vanishes) to obtain
\bea
K_L&=&K_g+Z_1T+\bar Z_1\bar T+Z_2(S+iP)+\bar Z_2(S-iP)\cr
\cr
&=&K_g+\phi(V_2-V_1)+\bar\phi(\bar V_2-\bar V_1)+\j(V_2-\bar V_1)+\bar\j(\bar V_2-V_1),
\label{legendre}
\eea
where in the second step we substituted the superfield strengths in terms of the prepotentials, and integrated by parts twice. Therefore, $\phi$ is chiral and 
$\j$ is twisted chiral.  Lastly, because the prepotential gauge transformation is a shift by a semichiral superfield we can choose the gauge in which $X=0$ and 
$Y=0$. This yields the duality functional
\bea
\label{dfunc}
K_D&=&K(i(V_1-V_2),i(\bar V_2-\bar V_1),i(V_1-\bar V_1),-i(V_2-\bar V_2))\cr
\cr
&~&+\phi(V_2-V_1)+\bar\phi(\bar V_2-\bar V_1)+\j(V_2-\bar V_1)+\bar\j(\bar V_2-V_1)\cr
&&\mbox{or}\cr
K_D&=&K(i(V_2-\bar V_1),i(V_1-\bar V_2),i(V_1-\bar V_1),-i(V_2-\bar V_2))\cr
\cr
&~&+\phi(V_2-V_1)+\bar\phi(\bar V_2-\bar V_1)+\j(V_2-\bar V_1)+\bar\j(\bar V_2-V_1)
\eea
To see how we recover the original K\"ahler potential we study the 
constraints imposed by the Lagrange multipliers.  
The $\phi$ and $\j$ equations of motion require
\bea
V_2-V_1&=&iX+iY\cr
\bar V_2-V_1&=&i X+i\bar Y.
\eea
Plugging this back into (\ref{dfunc}) we obtain the original potential.

If on the other hand, we impose the equations of motion of the prepotentials, 
solve for $V_1$ and $V_2$ and substitute back into (\ref{dfunc}), we
obtain the T-dual K\"ahler potential.  
This duality replaces a pair of left and right semichiral superfields 
with a pair of chiral and twisted chiral superfields.  

We would like to mention that the duality functional obtained 
before appears to be 
related to the Legendre transforms described in \cite{gmst}. The
authors of \cite{gmst} began by writing the K\"ahler potential as 
$K=K(V,\bar V,W,\bar W)-(XV+YW+c.c)$, where $X,Y$ are left, right semichiral superfields and $V,W$ are unrestricted. If the K\"ahler potential has an isometry, 
resulting in a dependence of only three real independent linear 
combinations of the unconstrained complex $V$ and $W$, then
by integrating out the semichiral superfields, one is left with a K\"ahler 
potential expressed in terms of chiral and twisted chiral superfields.

As a concrete example of the T-duality map we consider the torus $T^4$. Its
(2,2) supersymmetric sigma model action is derived from the K\"ahler potential
\be
K=R(\bar X+Y)(X+\bar Y)-\frac{R}{4}(\bar Y+Y)^2.
\ee
The duality functional is
\be
K_D=R(V_2-\bar V_1)(\bar V_2-V_1)+\frac{R}{4}(V_2-\bar V_2)^2+\phi(V_2-V_1)+
\bar\phi(\bar V_2-\bar V_1)+\j(V_2-\bar V_1)+\bar\j(\bar V_2-V_1).
\ee
The dual potential, up to generalized K\"ahler gauge transformations, reads
\be
\tilde K=\frac{1}{R}(\bar\phi\phi-\bar\j\j).
\ee
This is indeed is the potential for the T-dual $T^4$, 
this time written in terms of 
chiral and twisted chiral superfields. As expected, the radius $R$ 
of the dualized $S^1$ is mapped into $1/R$.

\subsection{T-duality and the Killing potentials}

Having identified the Killing potentials of the theory in terms of the K\"ahler potential, we would like to highlight their role in the T-duality map.   
Let us recall the more familiar situation encountered when T-dualizing  along 
a chiral superfield direction, and   consider the  U(1) Killing vector: 
\be
\x=\x^i\del_i+\bar\x^{\bar i}\del_{\bar i}.
\ee
{}From the invariance of the K\"ahler potential $\x K=0$ we derive that
\be
\x^i\del_i K=-\bar\x^{\bar i}\del_{\bar i}K.
\ee
This implies that $\mu(\phi,\bar\phi)=-i\x^i\del_iK$ is the Killing 
potential. $\mu$ is also the moment map \cite{hklr}.  Next we turn to the T-duality 
functional \cite{rv} which is constructed starting from the 
gauged K\"ahler potential \cite{hullklr}: 
\be
K_g=K(\tilde\phi,\phi,Z^a),
\ee
where $\tilde \phi=e^{-iV\bar\x}\bar \phi$. 
This means that $K_g$ has the exact same dependence on $(\phi,\tilde\phi)$ as $K$ has on $(\phi,\bar\phi)$.  By varying $K_g$ with respect to $V$ we get
\be
\frac{\del K_g}{\del V}=\frac{\del \tilde\phi^{\bar i}}{\del V}\frac{\del K_g}{\del \tilde\phi^{\bar i}}=(-i\bar\x\tilde\phi^{\bar i})\frac{\del K_g}{\del \tilde\phi^{\bar i}}=-i\tilde\x^{\bar i}\frac{\del K_g}{\del \tilde\phi^{\bar i}}=\mu(\phi,\tilde\phi)\equiv\tilde\mu,
\ee
where $\tilde\x^{\bar i}=e^{-iV\bar\x}\bar\x^{\bar i}$.  As explained previously, the duality functional is given by the gauged K\"ahler potential plus the 
Lagrange multipliers enforcing the pure gauge condition
\be
K_D=K_g-(\bar\j+\j)V
\ee
with $\j$ a twisted chiral superfield. 
The T-dual potential is obtained by imposing the prepotential equation of motion
\be
\frac{\del K_D}{\del V}=\frac{\del K_g}{\del V}-(\bar\j+\j)=\tilde \mu-(\bar\j+\j)=0.
\ee
So, T-duality embeds the moment map $\tilde\mu$ as the real part of the dual coordinate. 

 A similar story goes through when considering sigma models with semichiral superfields.  Starting with a potential
$
K=K(\bar X,X,\bar Y,Y)
$ which is invariant under the action of the isometry generator $\xi$, $\xi K=0$, we can derive the same equations as those given above.  
In this case, however, there are three Killing potentials instead of one
\bea
\x^i\del_iK+\x^{i'}\del_{i'}K&=&-\bar\x^{\bar i}\del_{\bar i}K-\bar\x^{\bar i'}\del_{\bar i'}K=i\mu_1(\bar X,X,\bar Y,Y)\cr
\cr
\x^i\del_iK+\bar\x^{\bar i'}\del_{\bar i'}K&=&-\bar\x^{\bar i}\del_{\bar i}K-\x^{i'}\del_{i'}K=i\mu_2(\bar X,X,\bar Y,Y)\cr
\cr
\x^i\del_iK+\bar\x^{\bar i}\del_{\bar i}K&=&-\x^{i'}\del_{i'}K-\bar\x^{\bar i'}\del_{\bar i'}K=\s(\bar X,X,\bar Y,Y).
\eea

The potential is gauged by making the substitutions in (\ref{replacement}):
$K_g=K(\bar{\tilde X},\tilde X,\bar{\tilde X},\tilde Y)$.
Just as above, we note that
\bea
\frac{\del K_g}{\del V_1}&=&\frac{\del \tilde X^i}{\del V_1}\frac{\del K_g}{\del \tilde X^i}= \tilde\x^i\tilde\del_iK_g\cr
\cr
\frac{\del K_g}{\del V_2}&=&\frac{\del \tilde Y^{i'}}{\del V_1}\frac{\del K_g}{\del \tilde Y^{i'}}= \tilde\x^{i'}\tilde\del_{i'}K_g.
\eea

With the duality functional given by (\ref{legendre}), 
the complex prepotential equations of motion are
\bea
&&\frac{\del K_D}{\del V_1}=\frac{\del K_g}{\del V_1}+\f+\bar\j=\tilde\x^i\tilde\del_iK_g+\f+\bar\j=0\nn\\
&&\frac{\del K_D}{\del V_2}=\frac{\del K_g}{\del V_2}-\f-\j=\tilde\x^{i'}
\tilde\del_{i'}K_g-\f-\j=0.
\eea
After rewriting them as
\bea
-\tilde\x^i\tilde\del_iK_g&=&\f+\bar\j\cr
\cr
-\bar{\tilde\x}{}^{\bar i}\tilde\del_{\bar i}K_g&=&\bar\f+\j\cr
\cr
-\tilde\x^{i'}\tilde\del_{i'}K&=&-\f-\j\cr
\cr
-\bar{\tilde\x}{}^{\bar i'}\tilde\del_{\bar i'}K&=&-\bar\f-\bar\j,
\eea
it is clear that only three of the four equations of motion are 
independent, since
\be
\tilde\x^i\tilde\del_iK_g+\bar{\tilde\x}{}^{\bar i}\tilde\del_{\bar i}K_g+\tilde\x^{i'}\tilde\del_{i'}K_g+\bar{\tilde\x}{}^{\bar i'}\tilde\del_{\bar i'}K_g=0.
\ee
This is, of course, the statement that the K\"ahler potential possesses an 
isometry. The content of the remaning three equation is as follows:

-the Killing potential $\tilde \mu_1$ is mapped to the imaginary part of $\j$:
\be
\tilde\x^i\tilde\del_iK_g+\tilde\x^{i'}\tilde\del_{i'}K_g=i\mu_1(\bar{\tilde X},\tilde X,\bar{\tilde Y},\tilde Y)=\j-\bar\j,
\ee
-$\tilde\mu_2$ is mapped to the imaginary part of $\f$:
\be
\tilde\x^i\tilde\del_iK_g+\bar{\tilde\x}{}^{\bar i'}\tilde\del_{\bar i'}K_g=
i\mu_2(\bar{\tilde X},\tilde X,\bar{\tilde Y},\tilde Y)=\bar\f-\f,
\ee
-the third Killing potential $\tilde\s$ is mapped to the sum of the 
real parts of $\f$ and $\j$:
\be
\tilde\x^i\tilde\del_iK_g+\bar{\tilde\x}{}^{\bar i}\tilde\del_{\bar i}K_g=\s(\bar{\tilde X},\tilde X,\bar{\tilde Y},\tilde Y)=-\f-\bar\f-\j-\bar\j.
\ee

\section{Quotients}

Lastly we address the quotient construction for the semichiral sigma 
models. The quotient manifold is obtained by extremizing the gauged K\"ahler 
potential with respect to the
three real linear independent combinations of the prepotentials for each of 
the isometry group generators. The dimension
of the quotient manifold, which remains bi-hermitean, is 
$dim\,{\cal M}-4dim \,G$.

After adding the FI terms\footnote{The quotient construction relies on the same duality functional as used for the T-duality map. The FI terms correspond to the 
Lagrange multiplier terms, where the Lagrange multipliers are taken to be 
constant.} 
\be
{\rm FI\; terms}=
\D K_g=r_0\tilde V+r_1\tilde V_1+r_2\tilde V_2
\ee
where
\bea
\tilde V&=&Re(V_2)-Re(V_1)\cr
\tilde V_1&=&Im(V_1)\cr
\tilde V_2&=&Im(V_2)
\eea
to the gauged K\"ahler potential 
given in (\ref{kgmu}), and extremizing with respect to $\tilde V,\tilde V_1$ 
and $\tilde V_2$, one finds
\be
e^{2i\tilde V_1\xi_1}(\mu_1+\mu_2)=r_1, 
\qquad e^{2i \tilde V_2 \xi_2}(\mu_1-\mu_2)=r_2,
\qquad e^{2\tilde V\xi_3}\sigma=r_0.
\ee 
We end with three constraints, corresponding to the three Killing potentials
that can be defined when gauging the semichiral sigma model.  
These are the equations which define the quotient. In practical terms, one
solves them for the prepotentials, and substitutes back into the gauged 
K\"ahler potential to arrive at the quotient manifold potential. 
We recall 
that a similar constraint (involving the only Killing potential) 
defines the K\"ahler quotient on a K\"ahler manifold \cite{hklr}. 

Let us consider the the flat space quotient as an example.  
We will quotient a rotation isometry instead of a shift isometry.  
We take two copies of $R^4$ and gauge the fields for both copies with 
the same charge (which we have set to 1).  The fields $X_1$ and $Y_1$ will belong to the first $R^4$ and $X_2$ and $Y_2$ to the second.  The global $U(1)$ phase transformation will take the form $X\ra e^{i\e}X$, $Y\ra e^{-i\e}Y$ with $\e$ a real constant.  Under the local gauge transformations the fields will transform as
\be
\label{gaugetrans}
X_{1/2}\ra X'_{1/2}=e^{i\L}X_{1/2},~~~ Y_{1/2}\ra Y'_{1/2}=e^{-iU}Y_{1/2}~~.
\ee
The gauged action is
\bea
K_g&=&e^{i(V_1-\bar V_1)}(\bar X_1 X_1+\bar X_2 X_2)+ e^{i(V_1-V_2)}(Y_1X_1+Y_2X_2)
+e^{i(\bar V_1-\bar V_2)}(\bar Y_1\bar X_1+\bar Y_2\bar X_2)\cr
&~&+\frac{1}{2}e^{i(\bar V_2-V_2)}(\bar Y_1 Y_1+\bar Y_2 Y_2)+\D K_g
\eea
The equations of motion for the prepotentials read
\bea
&&
-2(\bar X_1 X_1+\bar X_2 X_2)e^{-2\tilde V_1}-(Y_1X_1+Y_2X_2)e^{-i\tilde V}e^{-\tilde V_1}e^{\tilde V_2}-(\bar Y_1\bar X_1+\bar Y_2\bar X_2)e^{i\tilde V}e^{-\tilde V_1}e^{\tilde V_2}=r_1\nn\\
&&(Y_1X_1+Y_2X_2)e^{-i\tilde V}e^{-\tilde V_1}e^{\tilde V_2}+(\bar Y_1\bar X_1+\bar Y_2\bar X_2)e^{i\tilde V}e^{-\tilde V_1}e^{\tilde V_2}+(\bar Y_1 Y_1+\bar Y_2 Y_2)e^{2\tilde V_2}=r_2\nn\\
&&-i(Y_1X_1+Y_2X_2)e^{-i\tilde V}e^{-\tilde V_1}e^{\tilde V_2}+i(\bar Y_1\bar X_1+\bar Y_2\bar X_2)e^{i\tilde V}e^{-\tilde V_1}e^{\tilde V_2}=r_0
\eea
To solve these equations we introduce the notation: 
$x=(\bar X_1 X_1 + \bar X_2 X_2), y=(\bar Y_1 Y_1 + \bar Y_2 Y_2), 
z=(Y_1X_1+Y_2X_2),
A=(Y_1X_1+Y_2X_2)e^{-i\tilde V}e^{-\tilde V_1}e^{\tilde V_2}$.
Then we can rewrite them as
\be
2 Im(A)=r_0, \qquad y e^{2 \tilde V_2} + 2 Re A= r_1, -2x e^{-2\tilde V_1}+
y e^{2\tilde V_2}=r_1+r_2.  
\ee
Next, solving for $|A|$ we find
\be
|A|^2=z\bar z e^{-2\tilde V_1+2\tilde V_2}=\frac{r_0{}^2}4+\frac 14
(r_2-y e^{2\tilde V_2})^2
\ee
Further substituting $\tilde V_1$ in terms of $\tilde V_2$ yields
\be
z\bar z e^{2\tilde V_2}(y e^{2\tilde V_2}-r_1-r_2)=\frac{x}2(r_0{}^2
+(r_2-y e^{2\tilde V_2})^2
\ee
which may be solved directly for $\tilde V_2$, giving
\be
e^{2\tilde V_2}=\frac{(r_1+r_2)\bar zz-r_2xy \pm \sqrt{((r_1+r_2)\bar zz-r_2xy)^2+2(r_2^2+r_0^2)(\bar zz-\frac{1}{2}xy)xy}}{2y(\bar zz-\frac{1}{2}xy)}
\ee
The reality of $\tilde V_2$ will require that $\bar zz\geq \frac{1}{2}xy$ indicating the presence of a boundary in the quotient target space.  The solutions for $\tilde V_1$ and $\tilde V$ follow with
\bea
e^{-2\tilde V_1}&=&\frac{1}{2x}(ye^{2\tilde V_2}-r_1-r_2)\cr
\cr
e^{-i\tilde V}&=&\frac{1}{2z}e^{\tilde V_1}e^{-\tilde V_2}(r_2+ir_0-ye^{2\tilde V_2})~~.
\eea
To complete the discussion of the quotient, we have to choose a gauge.  Considering (\ref{gaugetrans}), we will pick the gauge where $X'_1=1$, $X'_2=\frac{X_2}{X_1}$, $Y'_1=1$ and $Y'_2=\frac{Y_2}{Y_1}$.  Despite the complexity of the final answer, the gauge fixing step demonstrates that the dimension of the quotient manifold is smaller by 4 (which was the expected result since we quotient a U(1) isometry).   Bi-hermiticiy of the quotient geometry is guaranteed since $(2,2)$ supersymmetry has been preserved.  An interesting point is that since the quotient potential is more than quadratic in the fields, the quotient target space has non-trivial H flux.
\section*{Acknowledgments}

We are grateful to  L.~Pando Zayas for collaboration during an early 
stage of this work. We are also thankful to S.J.~Gates for useful discussions.
This work is  partially supported by DOE.

As we were completing our work, we became aware of
related work by U.~Lindstrom, M.~Rocek, I.~Ryb, R.~von Unge and M.~Zabzine; 
we thank them for agreeing
to delay their work and post simultaneously.

\end{document}